# Reduced neural activity during volatile anesthesia compared to TIVA: evidence from a novel EEG signal processing analysis

**Amitai Bickel** [1,2], **Neta B. Maimon** [3,4] **Lior Molcho** [4], **Nathan Intrator** [4,5], **Shimon Ivry** [6], <u>**Alexey**</u> **Gavrilov**[6]

[1] Department of Surgery A, Galilee Medical Center, Nahariya, Israel

[2] Faculty of Medicine in the Galilee, Bar-Ilan University, Safed, Israel.

[3] The School of Psychological Sciences, Tel Aviv University, Tel Aviv, Israel.

[4] Neurosteer LTD, Herzliya, Israel.

[5] Blavatnik School of Computer Science and Sagol School of Neuroscience, Tel Aviv University, Tel Aviv, Israel.

[6] Neurosteer Inc, New York, NY, USA

[6] Department of Anesthesiology, Galilee Medical Center, Nahariya,

**\* Correspondence:**

Neta B. Maimon

netacoh3@mail.tau.ac.il

## Abstract

**Background:** Post-operative cognitive dysfunction/ decline is a well-known phenomenon and of crucial importance especially in the elderly, emphasizing the significance of selecting the proper anesthetics. General anesthesia can be accomplished by inhalation-based (volatile) or total intravenous anesthesia (TIVA). While their effects on post-operative symptoms have been investigated, little is known about their influence on brain functionalities during the surgery itself.

**Objective:** To assess differences in brain activity between volatile and TIVA anesthetics during surgery.

**Methods:** Seventeen patients who were electively scheduled for laparoscopic cholecystectomy in Galilee Medical Center, Nahariya, gave written informed consent to participate in the study, and were randomly divided to receive either volatile anesthesia (n=9), or TIVA (n=8). The level of anesthesia was kept to be equal in both groups. A single bipolar EEG electrode (Aurora by Neurosteer®) was placed on the participants' foreheads. It presented real-time activity and collected their data during the surgery. The dependent variables included frequency bands (delta, theta, alpha, and beta), and three features (VC9, ST4, and A0) previously extracted with the Aurora device and provided by Neurosteer®.

**Results:** All surgeries were uneventful, and all patients showed bispectral index (BIS) score less than 60. Feature activity under volatile anesthesia (in comparison to TIVA) was significantly lower for the delta, theta and alpha frequency bands and for the three features (VC9, ST4 and A0). Further analysis showed that the largest difference between anesthesia types was for feature A0.

**Conclusions:** Both EEG frequency bands and novel brain activity features provide evidence that volatile anesthesia further reduces components of brain activity in comparison to TIVA anesthesia. Specifically, A0, which previously showed a correlation with cognitive decline severity and cognitive load, exhibited the most prominent difference between anesthesia types. Together, this study suggests that measuring brain activity during anesthesia using sensitive features, enables revealing that different anesthesia types may affect brain activity differently, which could affect the recovery from anesthesia, and consequently reduce post-operative cognitive decline.

# 1    Introduction

Over 50 million inpatient operations are performed in the US each year. Currently, general anesthesia by either total intravenous anesthesia (TIVA) or the traditional inhalation-based anesthesia (volatile) enables the anesthesiologist to gain desirable hypnotic and analgesic effects. The anesthetic dose is adapted to each patient's needs in a specific clinical situation. However, at present, there is a paucity of information regarding the influence of diverse modes of anesthetics on brain function (1,2).

The research investigating the differences between the two anesthetics include studies with diverse endpoints applying measures such as physiologic outcomes and postoperative effects (3-5). Based on clinical and basic research studies it seems that volatile anesthesia has more significant adverse cerebral effects in comparison to TIVA (6-9).

Postoperative cognitive dysfunction as manifested with a decline in brain function is a well-known non-specific anesthesia-associated phenomenon (1, 10). Moreover, studies in animals have demonstrated postoperative cerebral pathology, mainly following inhalational anesthetics (11-16). However, no study has been done to compare the sensing of neurological activation during surgery under the influence of volatile and TIVA.

Volatile anesthetics involving sevoflurane may cause myocardial contractility depression through direct negative inotropic effect upon the left ventricle and atria, in addition to left ventricular diastolic dysfunction (17). Volatile anesthetics may also cause chronotropic injury through sino-atrial depression. Such mechanisms are complex, depending on dosage, drugs and interactions. The intra-cellular effect is attributed to changes in the intra-cellular transport of calcium in the myocardium. Nitrous oxide may have a direct myocardial inotropic effect as well (18,19).

Using TIVA (as was used in our study) the propofol directly depresses the myocardium and causes decreased systemic vascular resistance. It also affects the autonomic sympathetic nervous system and has a direct effect on the vascular smooth muscle, causing arterial and venous vasodilatation, which contribute to blood pressure reduction. By affecting the baroreceptor mechanism, mild increase in heart rate ensues (18, 20). Nevertheless, the cardio-depressive effects of propofol can be controlled by careful titration and with drug interaction with short-acting opiates such as remifentanil. In general, low dose opiates induce minimal cardiovascular effects, and their main use is to induce analgesia (together with anesthetic drugs) to improve anesthesia (21).

Therefore, different brain activity under TIVA compared to volatile, should not stem from cardiovascular reasons, due to significant cerebral perfusion auto-regulation.

Recently, significant medical and scientific computerized progress has enabled better knowledge and comparison of the influence of both techniques of anesthesia on brain functionality (22, 23). Two main scientific developments have recently enabled the comparison of different anesthetic modalities' influence on the brain. First, it is now possible to measure the pharmacokinetics through respiratory gas monitoring. This indicates how volatile aesthetics gain access to the circulation indirectly through the lungs and enables accurate administration of the inhaled drug to a desired concentration (pharmacokinetic exactness). Measurement of minimum alveolar concentration reflects the pharmacodynamic exactness (24).

The major advances in total intravenous anesthetics stem from the advent of new drugs (25). Propofol, with appropriate pharmacokinetic (PK) and pharmacodynamic (PD) properties, is a short acting drug with a relatively quick nausea-free recovery (26,27). In addition, using an esterase-metabolized opioid, remifentanil, with a very high clearance, enables rapid achievement of a steady state after beginning its infusion, as well as recovery from TIVA (28,29). Using computerized algorithm and electronic pump, the concept of target-controlled infusion (TCI) systems, based on a pharmacokinetic-model algorithm, calculates the necessary infusion rates to achieve the targeted concentration (25,30,31). Based on high-resolution PK/PD models, unique advisory systems bring relevant information during anesthesia to improve pharmacologic control. Computer simulation helps in clinical application of PD synergistic interaction of propofol and remifentanil (32,33). All the above legitimizes the comparison of TIVA with volatile anesthetics, as addresses in our study.

Recent advances in EEG data analysis may be of use in the comparison of different anesthetic modalities. Spectral analysis, is the most used method to decompose the EEG signal, can summarize the brain activity states into activity levels of approximately eight frequency bands, each on a different location of the skull, using a multi-electrode setup. This approach provides an association between various brain states or disorders and EEG states in different locations (34, 35). Specifically, and relevant to our research, the delta frequency band has been associated with different sleep stages and wakefulness states and might be a good indicator in measuring differences between anesthetics (36-37).

In the present study, in addition to the classical spectral analysis, we applied an advanced hierarchical time-frequency analysis creating a novel EEG decomposition method. This method produces many brain-activity features (BAFs) (38). The processed data is presented on a two-dimensional information matrix, which can translate brain activity into a heat color map, to represent the magnitude of different BAFs in time. Next, for data analysis, higher-order features which were previously extracted (39) are provided with a second-by-second activity. These previously shown to correlate with cognitive load (40-42) and with cognitive decline severity (43).

The present study goal was to compare the brain activity under TIVA and volatile anesthesia. It was done using the single-channel EEG device: Aurora by Neurosteer®, which provided real-time presentation of the BAFs during the surgery, and frequency bands and novel features activity for further data analysis. To meet this goal, we ran a pilot study with 17 patients who underwent general anesthesia laparoscopic surgeries. Anesthesia depth of the patients (nine volatile anesthesia and eight TIVA) was assessed by both BIS (44-45) and the Aurora device. Differences in brain activity under the two anesthesia types may unveil potential effects on the recovery from general anesthesia.

## 2    Materials and Methods

Ethical approval for this study was provided by the Ethical Committee Institutional Helsinki committee of Galilee Medical Center, Nahariya, Israel (Ethical Committee N° 0137-13-NHR). In the beginning, the previous study was meant to test two modes of ventilation during laparoscopic cholecystectomy, conventional Vs. high frequency jet ventilation (HFJV). As HFJV should be combined with total intra-venous anesthesia (TIVA), all the study subjects gave informed consent to both the modes of ventilation and anesthesia as well. Our current study was later derived from the former one (methods, relevant results). The primary goal of our current study was to compare two modes of anesthesia.

### 2.1    Patients

Our study included 17 patients who were electively scheduled for laparoscopic cholecystectomy due to symptomatic cholecystolithiasis. All patients signed a written informed consent to participate in the experiment. All surgeries occurred at the Galilee Medical Center, Nahariya, Israel.

Mean age was 45.7±14.63 years (range 23-73, only one patient was aged 73, and age distribution in both cases was the same in both groups), M/F ratio 6/11. All patients were in a good medical state, with ASA score I-II. Pre-operative blood tests including renal and liver function tests were within normal range. All participants did not take any regular medication, and their cognitive status was normal. During surgery, patients were ventilated by conventional intermittent positive pressure (IPPV). The patients were randomly divided into two groups: those undergoing volatile anesthesia (9 patients) and those undergoing TIVA (8 patients). Randomization was achieved by choosing a sealed envelope containing the mode of anesthesia. Both groups were statistically matched in relation to their demographic and medical parameters.

## 2.2    Procedure

The main laparoscopic approach included abdominal $CO_2$ insufflation (digitally controlled) to 14 mmHg and introduction of 4 trocars/cannula into the peritoneal cavity. Following abdominal exploration via a video-camera, the cystic duct and vessels were identified, clipped, and divided, and the gallbladder dissected, resected and evacuated from the abdomen by an endo-bag. Surgery was terminated after peritoneal $CO_2$ evacuation. Mean duration of surgery was $38.2 \pm 8.46$ min. (range 25-55 min).

Anesthesia was administered by a senior anesthesiologist who is familiar with both TIVA and volatile anesthesia.  Monitoring of the depth of anesthesia was done by clinical assessment according to standard pathophysiological criteria, in conjunction with the BIS (44-45), keeping the BIS score between 40-60. When considering older subjects we kept the BIS index levels close to its higher level beside using physiologic parameters to assist in adjusting depth of anesthesia. Induction of anesthesia was the same in both anesthetized groups. Anesthesia was induced with intravenous administration of 1-3 mg dormicum (midazolam), 0.1 mg fentanyl, 150-200 mg propofol, and 0.5 mg/kg esmeron. The volatile anesthesia group received sevoflurane with nitrous oxide (60%) for maintenance. The concentration of sevoflurane ranged between 1.5% and 2.5%, making the adjustment according to the minimal alveolar concentration (MAC), hemodynamic parameters and BIS. For analgesia the patients under volatile anesthesia received fentanyl or remifentanil.
Patients in the TIVA group received 2-4 ng/ml remifentanil (46) and 2-4 µg/ml propofol (46,47) for maintenance, The titration of both drugs was done according to TCI regulations and the exact concentrations were affected by physiological parameters (clinical assessment) and BIS (47-51). For those drugs administration we used B Braun Perfusor Space Target Controlled infusion device pump (B Braun Medical Ltd. UK). Regarding the pump, we used the Schnider algorithm pharmacological model for propofol, and the Minto algorithm for remifentanil. adhering to TCI regulation and in accordance with BIS and basic clinical assessment. Esmeron was used in both treatment groups as needed for muscle relaxation.

The anesthesiologist attached the three electrodes on the patients' forehead. Real-time activity of the EEG features (described below) was presented during the surgery. For the data analysis, all participants' data was downloaded as second-by-second activity after all surgeries were done.

## 2.3 EEG Device

The EEG signal acquisition system included a three-electrode patch attached to the subject's forehead (Aurora by Neurosteer®., Herzliya, Israel). The medical-grade electrode patch included dry gel for optimal signal transduction. The electrodes were located at Fp1 and Fp2, with a reference electrode at Fpz. The EEG signal was amplified by a factor of 100 and sampled at 500 Hz. Signal processing was performed automatically by Neurosteer® in the cloud (see Section 2.1.4 below and Appendix A). We were therefore provided with a sample per second of activity of the brain oscillations (i.e., delta, theta, alpha, beta and gamma), and three selected features (i.e., VC9, ST4 and A0), see signal processing below.

## 2.4 Signal processing

Full technical specifications regarding the signal analysis are provided in Appendix A. In brief, the signal processing algorithm interprets the EEG data using a time/frequency wavelet-packet analysis, instead of the commonly used spectral or wavelet analysis. The Best Basis algorithm (52) constructs an orthogonal decomposition of the single EEG channel and together with the frequency bands creates a presentation of 121 features which we term brain activity features (BAFs). Unlike the standard frequency band features, the time/frequency wavelet packets include in addition to the time/varying components also higher harmonics.

To demonstrate this process, let $g$ and $h$ be a set of biorthogonal quadrature filters created from the filters G and H, respectively. These are convolution-decimation operators, where in a simple Haar wavelet, $g$ is a set of averages and $h$ is a set of differences.

Let $\psi_1$ be the mother wavelet associated with the filters $s \in H$, and $d \in G$. Then, the collection of wavelet packets $\psi_n$, is given by:

(1)

$$\psi_{2n} = H\psi_n; \qquad \psi_{2n}(t) = \sqrt{2}\sum_{j \in Z} s(j)\psi_n(2t - j),$$

(2)

$$\psi_{2n+1} = G\psi_n; \qquad \psi_{2n+1}(t) = \sqrt{2}\sum_{j \in Z} d(j)\psi_n(2t - j).$$

The recursive form provides a natural arrangement in the form of a binary tree. The functions $\psi_n$ have a fixed scale. A library of wavelet packets of any scale $s$, frequency $f$, and position $p$ is given by

(3)

$$\psi_{sfp}(t) = 2^{-s/2}\psi_f(2^{-s}t - p).$$

The wavelet packets $\{\psi_{sfp} : p \in Z\}$ include a large collection of potential orthonormal bases. An optimal basis can be chosen by the best-basis algorithm (52). Furthermore, an optimal mother wavelet can be chosen by (53). Following robust statistics methods to prune some of the basis functions, one gets 121 basis functions, which we term brain activity features (BAFs). Based on a given labeled-BAFs dataset, various models can be created for different discriminations of these labels. In the linear case, these models are of the form:

(4)

$$V_k(w, x) = \Psi\left(\sum_i w_i x_i\right),$$

where $w$ is a vector of weights and $\Psi$ is a transfer function that can either be linear, e.g., $\Psi(y) = y$, or sigmoidal for logistic regression $\Psi(y) = 1/(1 + e^{-y})$.

## 2.5   Dependent Variables

For details of creating the higher-level features VC9, A0 and ST4 see Appendix A and (39-41). Validation of the cognitive properties of these features on healthy individuals appears in (39,42) and usage in different context appears in (40,41,43).

We provide here a short description of the creation of the high-level features.

The 121 BAFs were created in an unsupervised way, namely no labels of brain status were used. The higher level features were created using linear and nonlinear machine learning techniques from previously collected labeled datasets. These datasets included EEG data collected from participants undergoing different cognitive and emotional challenges. Specifically, VC9 was found with linear discriminant analysis technique (LDA) on the 121 BAFs to be the best separator between an auditory detection task (higher cognitive load), and an auditory classification task (lower cognitive load, 43); A0 was calculated to discriminate between resting state task with music and auditory detection task using PCA; and ST4 was calculated to discriminate between high-load and low-load conditions in auditory n-back task, using unsupervised PCA.

On separate datasets, these features were found to be correlated with different tasks/levels/states. For example, VC9 correlated with cognitive load using the *n*-back task, auditory detection task, and interruptions paradigm (39-43). VC9 also correlated with cognitive load and individual performance of medical interns while undergoing a surgery simulation task (40). A0 and ST4 activity correlated with cognitive decline scores (i.e., mini-mental score, 54) and separated between groups with different levels of cognitive impairment (i.e., mild, moderate and severe) (39,43), ST4 correlated with individual performance tested on healthy participants undergoing the n-back task (39,41), and also showed some ability to differentiate between cognitive load levels (39,43). Neurosteer® provided the real-time activity of the BAFs and a dataset with a sample per second activity of the EEG bands (i.e., delta (0.5-4 Hz), theta (4–7 Hz), alpha (8–15 Hz), and beta (16–31 Hz)), and features (i.e., VC9, A0 and ST4).

## 2.6    Statistical analysis

Due to non-normal distribution of the data, we examined the differences (of median activities) between the two anesthetics using the Mann-Whitney test for continuous variables. For each comparison d prime was calculated to estimate effect size. Additionally, we compared the effect sizes of al dependent variables from smallest to highest to explore which of these effects was significantly larger than the others using the following:

If d is the observed Cohen's d value, then the sampling variance of d is approximately equal to:

$$v = \frac{1}{n1} + \frac{1}{n2} + \frac{d^2}{2(n1 + n2)},$$

So, to test $H_0 : \delta_1 = \delta_2$ (where $\delta_1$ and $\delta_2$ denote the true d values of the two studies), we computed:

$$z = \frac{d1 - d2}{\sqrt{v1 + v2}},$$

which follows approximately a standard normal distribution under *H0*. So, if |z|≥1.96, we rejected *H0* at α=.05 (two-sided).

The data was analysed using the These analyses were performed using Python Statsmodels (55) and SciPy (56).

## 3    Results

### 3.1    Real-Time Presentation

All surgeries were uneventful, without intra-operative hemodynamic instability. Post-operative convalescence was normal, and all patients were discharged on the first post-operative day. Renal and liver function tests remained within normal limits during the period of hospitalization. Depth of anesthesia was monitored by the regular techniques (including BIS) and showed no significant difference between the two anesthesia types. During surgery, all patients had a BIS reading below 60.

Figure 1 depicts brain activity of twelve patients from the current study, 6 during TIVA (A panels) and 6 during the volatile anesthesia (B panels) BAFs activity is shown in panel 1 and spectrogram activity is shown in panels 2 (1-30 Hz) and 3 (1-200 Hz). Both types of activity were generated using the same EEG data collected from the Aurora device. The BAF representation of the TIVA shows increased activity compared to the volatile anesthesia, The spectrograms panels do not exhibit this reduction.

The feature and band width results were extracted as an average over a period that started about 5 minutes after the initiation of anesthesia and ended right when the operation ended, just before anesthesia level started to be reduced.

The spectrogram representation is slightly distorted at the low frequencies due to the high pass filter used by the sensing front end. For this reason, the energy in delta is lower than expected. However, delta is still significantly different between volatile and TIVA anesthesia.

**A**

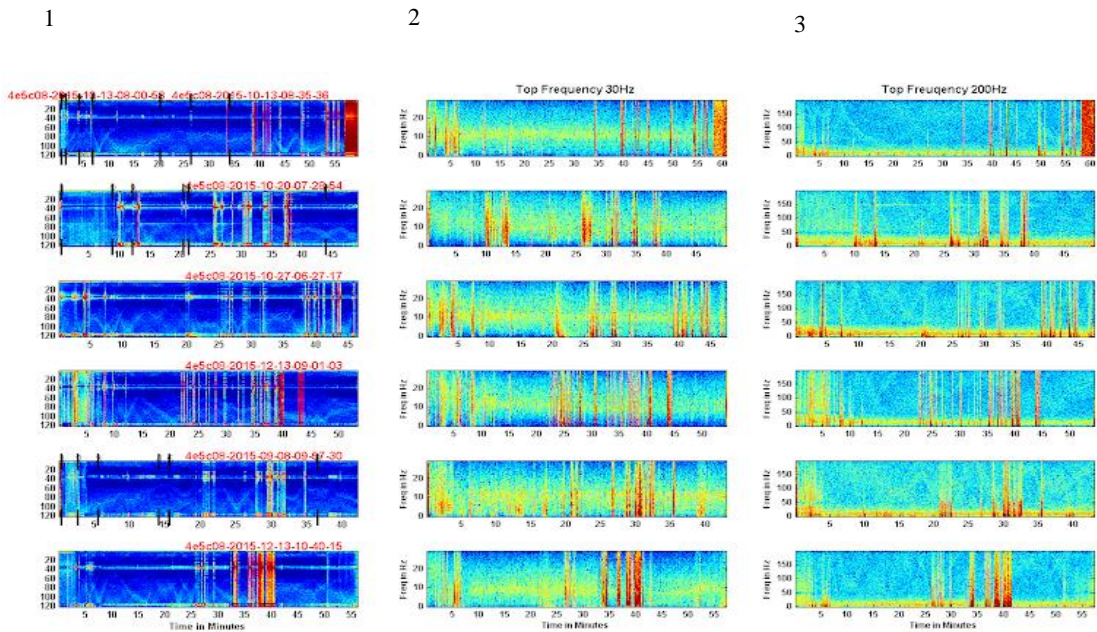

Figure 1- Comparison of real-time activity of 6 patients from the current study while undergoing TIVA (A) and 6 patients under volatile anesthesia (B), as a function of 121 BAFs (panel 1), frequency range 0-30 Hz (panel 2) and frequency range 0-200 Hz (panel 3).

## 3.2   Data Analysis Results

Averaged activity of each participant and feature or frequency bands are presented in figure 2. Following Mann-Whitney tests, the delta, theta, and alpha bands exhibited higher activity for TIVA than volatile ($p = 0.004$, $p = 0.007$, $p = 0.0151$ respectively), and beta band did not exhibit significant difference ($p = 0.096$). The VC9, ST4 and A0 features also showed a significant difference between the two anesthesia types ($p = 0.0151$, $p = 0.0375$, $p = 0.0003$, respectively). To explore which of the differences were larger, we calculated the effect size depicted by d prime and observed d prime of each comparison. Next, we compared the effect sizes to each other in ascending order. The highest effect size was found for the A0 feature, its effect was significantly larger than the second highest effect of the delta band ($p = 0.019$, see table 1 for all effects and comparisons).

A)

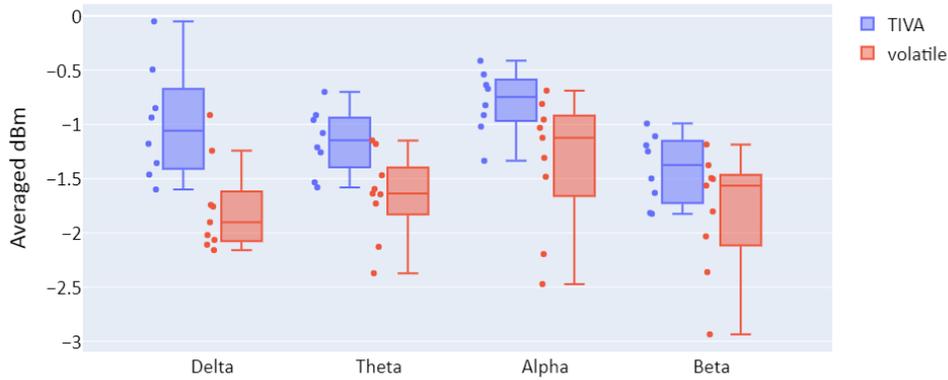

B)

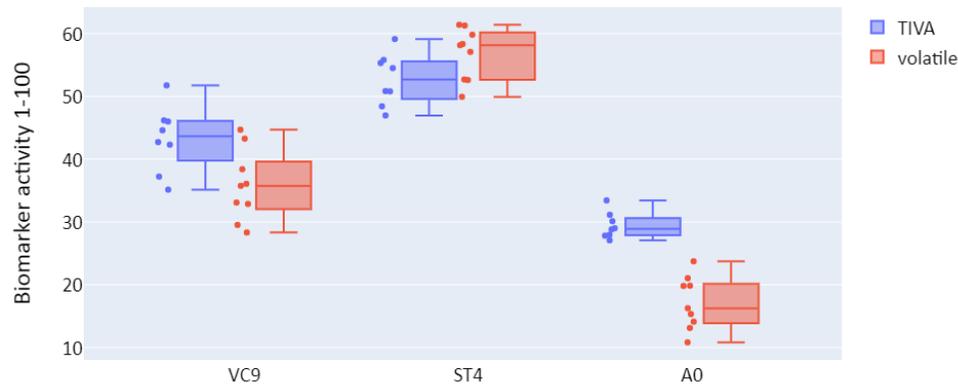

**Figure 2:** (A) The distribution of participant frequency bands activity (in dBm), and (B) The distribution of participants features activity (normalized between 1-100), as a function of anesthesia type: TIVA ($n = 8$, blue), volatile ($n = 9$, red).

| Feature | U value | P value | D prime | Observed D prime | Z value comparison | P value comparison |
|---------|---------|---------|---------|------------------|--------------------|--------------------|

| | | | | | |
|---|---|---|---|---|---|
| **Beta** | 22 | 0.097 | 0.864 | 0.259 | | |
| **ST4** | 17 | **0.038** | -0.991 | 0.267 | -0.174 | 0.431 |
| **Alpha** | 13 | **0.015** | 1.133 | 0.276 | -2.881 | **0.002** |
| **VC9** | 13 | **0.015** | 1.371 | 0.295 | -0.315 | 0.376 |
| **Theta** | 10 | **0.007** | 1.416 | 0.299 | -0.059 | 0.477 |
| **Delta** | 8 | **0.004** | 1.630 | 0.319 | -0.272 | 0.393 |
| **A0** | 0 | **0.000** | 3.697 | 0.663 | -2.086 | **0.019** |

**Table 1:** U value, *p* value, d prime, and the d sampling variance of each of the Mann-Whitney comparisons between TIVA and volatile anesthesia for each of the dependent variables (i.e., frequency bands and features), presented from lowest to highest effect size. compared z and p values represent the values of the effect size comparison between the current feature and the feature presented above.

## 4    Discussion

In this study, brain activity of patients during laparoscopic cholecystectomy, under volatile and TIVA anesthesia was measured using two analyses of a single-bipolar EEG signal. Although it may be suspected that different drugs and dosages in the volatile and TIVA groups might affect brain functionality differently (the difference derives merely from the different drugs that are prescribed during the maintenance period of anesthesia), we are convinced that both groups received equipotent drug dosages, as the depth of anesthesia in both groups was closely monitored and controlled by means of BIS (keeping BIS score between 60-40), and according to cardio-vascular parameters. Eventually, preliminary dosages would be changed accordingly. The issue of adjusting the level of anesthesia is of major importance, especially for the elderly patients that are more vulnerable to post-operative cognitive decline (post-operative delirium), and new studies still emerge to improve neuro-monitoring during anesthesia (66). In practice in our study, we kept BIS index levels close to its higher level (near 60) when considering older subjects, beside additional physiologic parameters to assist adjusting the depth of anesthesia, keeping in mind that there are some indications that BIS is less accurate in the elderly patients (67). This work may provide a different view- point to potential inaccuracies of current monitoring, suggesting that cognitive function indicators may be useful (39-43).   We did our best to adjust the level of anesthesia to be equal in both groups, making the independent parameter to be the different mode of anesthesia. Another factor that might affect the depth of anesthesia ~~is~~ is the residual propofol that was administered during the induction phase. Although this effect is not prolonged, it might have significance in relatively short duration of laparoscopic cholecystectomy. The influence of propofol in determining EEG pattern would be more significant for the volatile group in which we use sevoflurane, because both drugs share molecular (GABA receptors) and systems-level mechanisms to induce anesthetic effects (69). However, we assume that this effect is not of major significance as we did our best to keep the level of anesthesia equal in both groups.

Notwithstanding, the delta and the beta frequency bands showed significantly lower brain activity under volatile anesthesia compared to TIVA. Descriptively, the real-time presentation of the BAFs created with the wavelet-packet analysis decomposition, represented clear reduction in overall activity under volatile anesthesia compared to TIVA. The delta, theta, and alpha bands and all three features: VC9, ST4 and A0 exhibited reduced activity under volatile anesthesia relatively to TIVA, which was most prominent within the A0 feature.

In their article, Kim et al. (37) compared depths of anesthesia within the participant. They showed that slow delta wave levels (power) decreased in correlation to anesthesia depth and increased when consciousness regained. Additionally, delta frequency band levels (power) correlate with stages of volatile anesthesia depth (57). Following the claim that differences in brain activity between the two anesthetics exist, it is not surprising that the delta frequency band specifically showed significant decrease under volatile anesthesia. Moreover, from all frequency bands explored in the present study, the delta exhibited the most significant difference between the two anesthesia types.

Notwithstanding, the A0 feature showed the highest reduction in activity during volatile anesthesia, which was significantly larger than the delta band. This additional sensitivity of the high-leveled features was previously shown in previous research. A0 activity showed higher correlation with cognitive decline scores (i.e., mini-mental score, (54)) than all frequency bands and ST4 and VC9 features (39,43). Additionally, A0 separated between groups with different levels of cognitive impairment (i.e., mild, moderate and severe cognitive impairment) significantly better than the theta and delta bands (39-41,43). ST4 correlated with individual performance tested on healthy participants undergoing the n-back task (39,40), unlike any other frequency band or feature, and the VC9 exhibited higher sensitivity to lower cognitive-loads in healthy participants (i.e., when undergoing the *n*-back task, 39-40, and when undergoing tasks under the surgery simulator, 41).

In an attempt to explain the unexpected demonstration of enhanced brain activity (awareness pattern, as was expressed in the color map during TIVA), we should relate to at least two main aspects regarding volatile and TIVA anesthesia. First, the effect on the cardiovascular system (modulation of brain perfusion) and second, the effect of different anesthetics on the brain.

One hypothesis may be that volatile anesthetics can induce increase in intra-cranial pressure via cerebral vasodilation. Sevoflurane can decrease cerebral metabolic rate of oxygen consumption ($CMRO_2$) and depress EEG and brain activity at clinically tolerated doses (51). Despite that, increase in cerebral flow occurs ("uncoupling" phenomenon), which is the sum of indirect vasoconstrictor and direct vasodilating influences, and a diminished effect on cerebral autoregulation. However, sevoflurane relatively preserves autoregulation and is associated with less direct vasodilator effect. EEG changes during anesthesia are diverse and dose-dependent, and in addition to anesthesia are also induced by hypoxia, hypercarbia and hypothermia.

Second, exposure to inhalational anesthetics can cause neurotoxicity via activation of GABA receptors resulting in neuronal apoptosis (8,9). Furthermore, volatile anesthetics may antagonize the NMDA receptors to cause neuronal degeneration (8). Inhalation anesthesia was also found to modulate central nicotinic transmission, thus involving various neurocognitive dysfunctions (58). Studies using animal models have shown that inhalational anesthetics increase amyloid-$\beta$ accumulation in the brain, together with cytotoxicity and astrocytic gliosis (11-14, 59-60). There is an association between certain types of volatile anesthetics and neuronal production of pro-inflammatory cytokines like tumor necrosis factor and various interleukins, which lead to neuro-inflammation and POCD through diverse pathways (14). Inhalational anesthetics were also found to decrease synaptic neurotransmission by affecting function regulation (15,46). Inhalational anesthesia may alter intracellular calcium homeostasis, leading to neuro-degeneration and apoptosis, through excessive activation of inositol triphosphate receptors (16, 61). It should be stressed that these findings relate to diverse drugs. In this study, we used sevoflurane that is associated with pathogenesis of several cerebral diseases, but causes less apoptosis and calcium derangements than isoflurane (62).

Regarding the effects of intravenous anesthetics, most sedative-hypnotic drugs cause a proportional reduction in cerebral metabolism and blood flow, resulting in a decrease in intra-cranial pressure. Propofol probably has a cerebro-protective potential in cases of ischemia and hypoxia. Most intravenous hypnotics have similar EEG effects, depending on concentrations and drug interactions (63,64). In TIVA, propofol is combined with remifentanil (a strong short-acting opioid) to enable the optimal anesthetic titration. The differences found between anesthesia types may unveil a new reason for postoperative cognitive dysfunction (POCD), manifesting as a decline in brain function. This common phenomenon is linked to age, type of surgery, medical history and anesthesia (1,6). The mechanisms underlying the cognitive dysfunction following anesthesia exposure are still speculative. It seems that the role of volatile anesthesia is more significant than TIVA, and may involve dysregulation of excitatory neurotransmitters like N-methyl-D-aspartic acid (NMDA) and inhibitory neurotransmitters such as $\gamma$-aminobutyric acid (GABA) in the hippocampus (7).

To conclude, this study provides initial demonstration of the potential of novel specific cognitive features (39-43), suggesting that each mode of anesthesia affects the brain differently. Further research is required to further validate these findings and to determine whether volatile anesthesia can be better optimized given the potential of the cognitive features information. We did not, however, perform any cognitive tests as we did not expect those results, and all patients did not show any cognitive derangements whatsoever. Any functional change in our study could have been assessed qualitatively as well as quantitatively (with statistical significance). Currently, a strict explanation of our findings cannot yet be given. Both volatile and intravenous anesthetics modulate diverse CNS neurotransmitters, and perhaps the explanation rests on molecular grounds, such as diverse interactions through the α or the β subunits of the GABA receptor (65). Further research is needed to expand the use of the single-electrode EEG during general anesthesia. It should be stressed that the main limitation of the study is the small number of participants. However, the results, showing the difference between TIVA and volatile anesthesia were significant, and excepting one patient aged 73, the groups were similar regarding the relevant parameters, and adjustment of anesthetic medications to age was done. Nevertheless, the preliminary results of our study suggest and encourage further study to carefully test whether the cognitive features can be for neural monitoring during surgeries, and for the selection of the proper mode of anesthetics, which could potentially help preventing post-surgery cognitive decline, especially in the elderly.

## 6 Conflict of Interest

Author NI is employed by Neurosteer Inc, Authors LM and NM is employed by Neurosteer LTD. The remaining authors declare that the research was conducted in the absence of any commercial or financial relationships that could be construed as a potential conflict of interest.

## 7 Author Contributions

A.B. and N.I. conceived and planned the experiments. A.B., N.I., S.I. and A.G. carried out the experiments. N.I. and L.M. designed and provided tech support for the EEG device. N.I., N.M., and L.M. contributed to the interpretation of the results. A.B, N.I., and N.M. took the lead in writing the manuscript. All authors provided critical feedback and helped shape the research, analysis, and manuscript.

## 8 Acknowledgments

## 9 Tables

| Feature | U value | P value | D prime | Observed D prime | Z value comparison | P value comparison |
|---------|---------|---------|---------|------------------|--------------------|--------------------|
| **Beta** | 22 | 0.097 | 0.864 | 0.259 | | |
| **ST4** | 17 | **0.038** | -0.991 | 0.267 | -0.174 | 0.431 |
| **Alpha** | 13 | **0.015** | 1.133 | 0.276 | -2.881 | **0.002** |
| **VC9** | 13 | **0.015** | 1.371 | 0.295 | -0.315 | 0.376 |
| **Theta** | 10 | **0.007** | 1.416 | 0.299 | -0.059 | 0.477 |
| **Delta** | 8 | **0.004** | 1.630 | 0.319 | -0.272 | 0.393 |

| A0 | 0 | **0.000** | 3.697 | 0.663 | -2.086 | **0.019** |
|---|---|---|---|---|---|---|

**Table 1:** U value, *p* value, d prime, and the d sampling variance of each of the Mann-Whitney comparisons between TIVA and volatile anesthesia for each of the dependent variables (i.e., frequency bands and features), presented from lowest to highest effect size. compared z and p values represent the values of the effect size comparison between the current feature and the feature presented above.

## 10    Figures

A

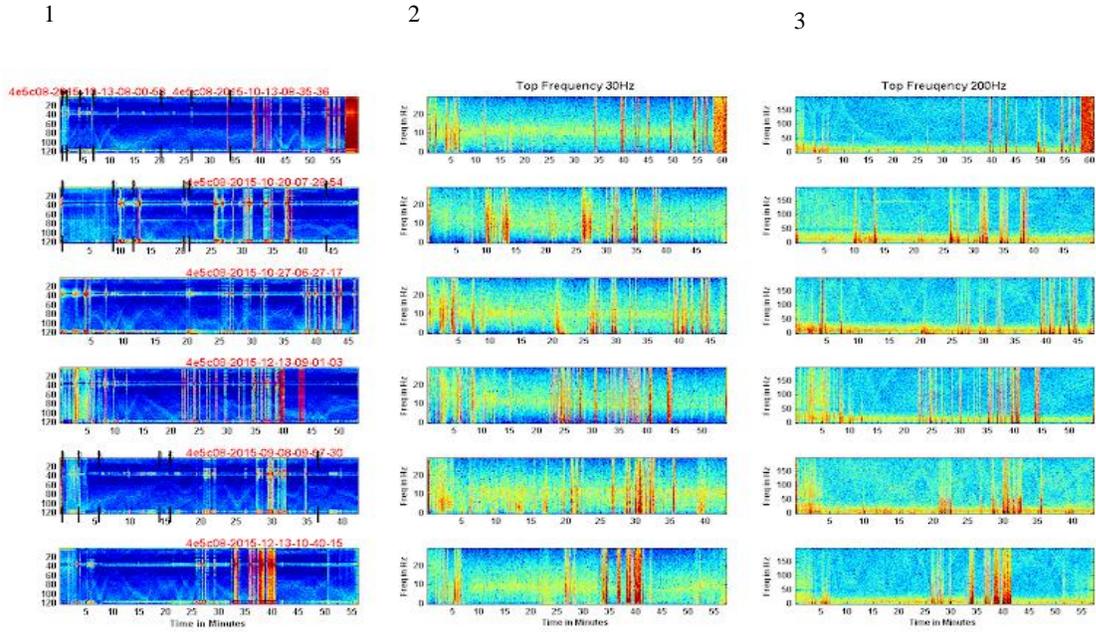

B

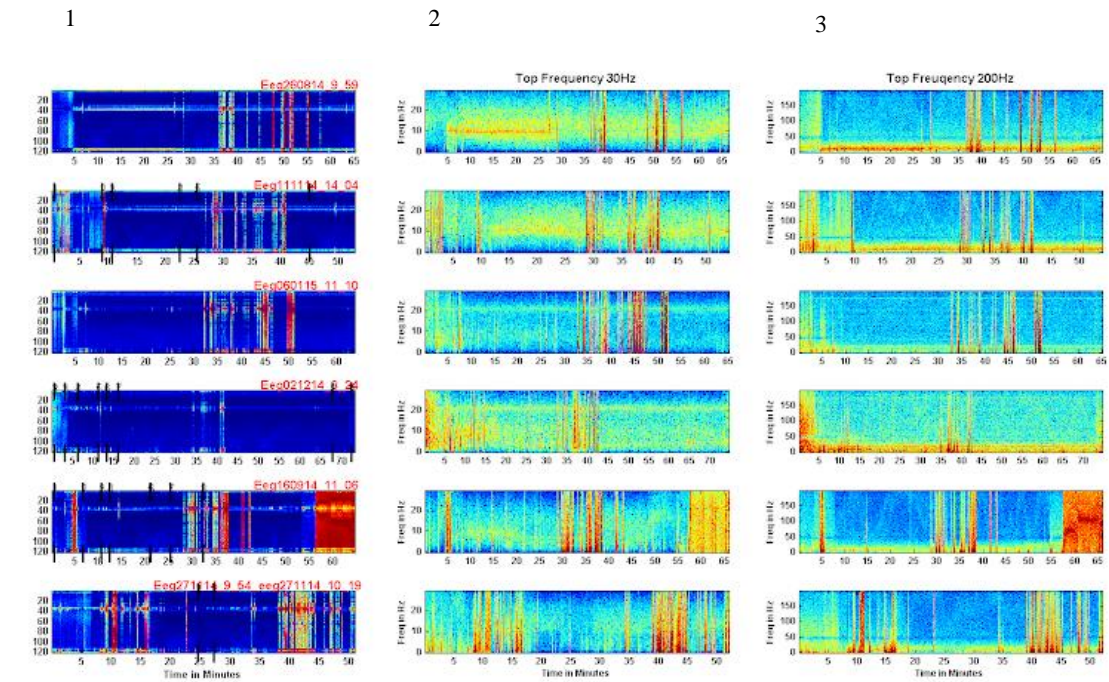

**Figure 1**- Comparison of real-time activity of 6 patients from the current study while undergoing TIVA (A) and 6 patients under volatile anesthesia (B), as a function of 121 BAFs (panel 1), frequency range 0-30 Hz (panel 2) and frequency range 0-200 Hz (panel 3).

A)

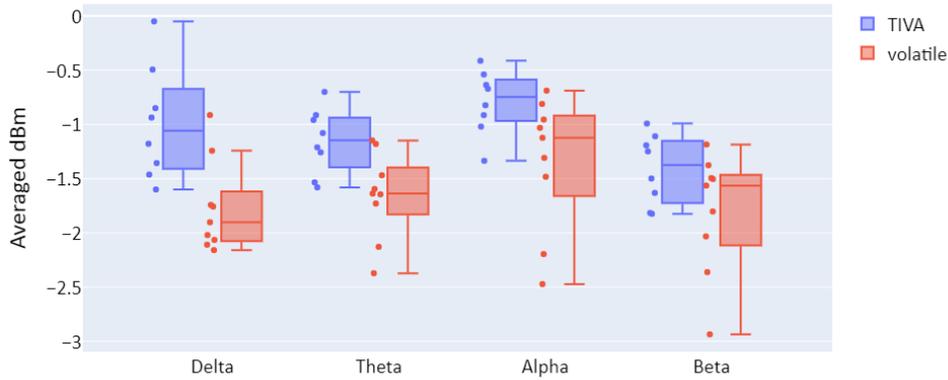

B)

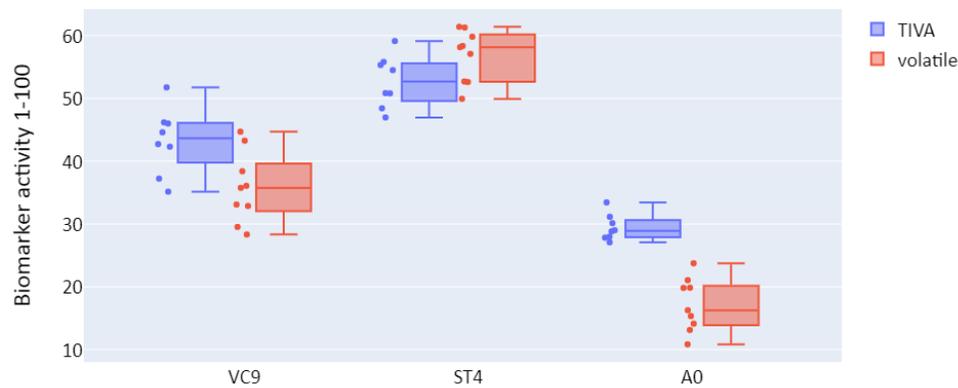

**Figure 2:** (A) The distribution of participant frequency bands activity (in dBm), and (B) The distribution of participants features activity (normalized between 1-100), as a function of anesthesia type: TIVA ($n = 8$, blue), volatile ($n = 9$, red).